\documentclass[pra,showpacs,amsmath,amssymb]{revtex4}

\usepackage{graphicx}
\usepackage{dcolumn}
\usepackage{bm}

\begin{document}

\preprint{}

\title{Photoassociation spectroscopy of cold calcium atoms}

\author{Carsten Degenhardt, Tomas Binnewies, Guido Wilpers, Uwe Sterr, Fritz Riehle}

\affiliation{Physikalisch-Technische Bundesanstalt, Bundesallee 100, 38116 Braunschweig, Germany}

\author{Christian Lisdat, Eberhard Tiemann}

\affiliation{Institut f\"ur Quantenoptik, Universit\"at Hannover, Welfengarten 1, 30167 Hannover, Germany}

\date{\today}
          
\begin{abstract}
Photoassociation spectroscopy experiments on $^{40}$Ca atoms close to the dissociation limit 4s4s$^1$S$_0-$4s4p$^1$P$_1$ are presented. The vibronic spectrum was measured for detunings of the photoassociation laser ranging from $0.6$~GHz to 68~GHz with respect to the atomic resonance. In contrast to previous measurements the rotational splitting of the vibrational lines was fully resolved. Full quantum mechanical numerical simulations of the photoassociation spectrum were performed which allowed us to put constraints on the possible range of the calcium scattering length to between 50~$a_0$ and 300~$a_0$.
\end{abstract}

\maketitle
\section{Introduction}
\begin{figure}
\centerline{\includegraphics[width=16cm]{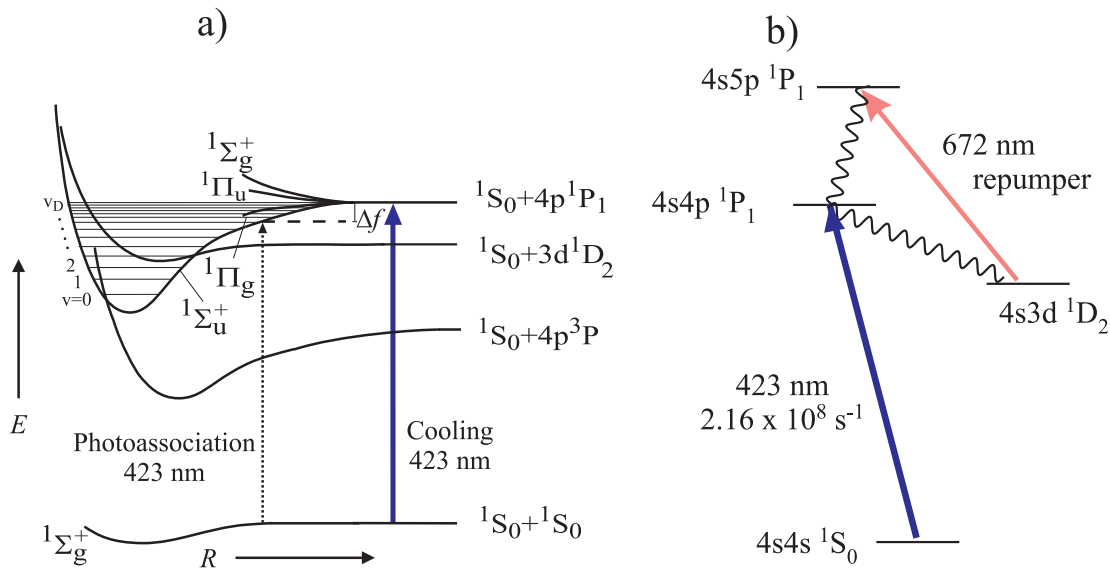}}
 \caption{Excerpt from the Ca molecular (a) and atomic (b) level scheme. The transition around $\lambda = 423$~nm is used for cooling and photoassociation. A repump laser ($\lambda = 672$~nm) prevents decay into metastable triplet levels via the {$^1D_2$} state. The long range part of the {$^1\Sigma_u^+$} molecular potential is probed in the experiment. The potential energies as a function of the internuclear distance $R$ are indicated schematically. $\Delta f$ is the frequency difference between the photoassociation laser and the atomic resonance.}
 \label{levels}
 \end{figure}

In photoassociation spectroscopy experiments within cold ensembles of atoms a bound molecule is formed under the influence of light. Such measurements can yield information about many atomic properties such as excited states lifetimes, ground state scattering lengths and long range potentials. Most of the experiments are performed with alkaline elements. However, due to their atomic fine and hyperfine structure theoretical calculations are in general very complicated to perform for these elements. In contrast, the most abundant isotopes of the alkaline earth metals show a non-degenerate ground state with no hyperfine splitting. This makes the comparison between experiment and theory easier than in the case of alkali metals. In Fig.~\ref{levels}a a simplified molecular level scheme is shown for the case of Ca$_2$. The photoassociation connects the ground state asymptote to the molecular state $^1\Sigma_u^+$. $\Delta f$ is the frequency difference between the photoassociation laser and the asymptote $^1$S$_0+4\textrm{p} ^1$P$_1$. Our first observations on photoassociation of Ca~\cite{zin00} were performed with cold ensembles around 3~mK and with short trap lifetimes of about 50~ms only, which in total leads to a poor signal-to-noise ratio for the trap loss by photoassociation. Only for the deepest recorded vibrational level at the excited asymptote we were able to observe rotational structure on the vibrational transition. The line profile and line position were interpreted by the help of assumptions, like an averaged fixed frequency shift of all transitions by 50~MHz due to 3~mK samples.

Here we will report of a more complete series of measurements under improved conditions, doubling the spectral interval of measured photoassociation lines, giving clear rotational structure and performing a full line profile simulation to get the proper shift of the line center due to thermal averaging and the intensity of the association laser.

\section{Experiment}
\begin{figure}
\centerline{\includegraphics[width=16cm]{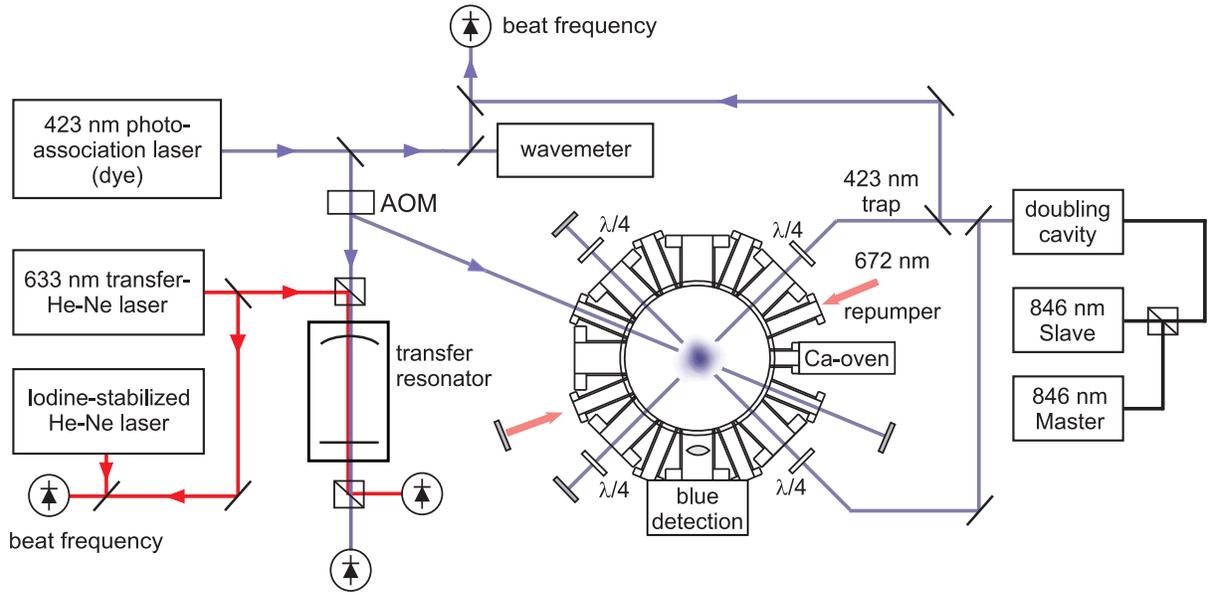}}
 \caption{The setup used in the experiments. Calcium atoms are captured from a thermal beam and cooled in a magnetooptical trap with three mutually orthogonal, retroreflected beams in $\sigma_+ - \sigma_-$ configuration (vertical beams not shown). The 423~nm trap light is generated by a frequency doubled master-slave diode laser combination. A dye laser produced the light used for photoassociation. For details see text.}
 \label{setup}
 \end{figure}

The calcium atoms are cooled and trapped in a magnetooptical trap (MOT) using the strongly allowed transition at $422.79$~nm ($\Gamma=2.16\cdot 10^8$ s$^{-1}$, see the level scheme in Fig.~\ref{levels}b). An overview of the experimental setup is shown in Fig.~\ref{setup}. The 423~nm radiation is generated by frequency doubling about 160~mW of a master-slave diode laser combination with a KNbO$_3$ crystal inside an enhancement cavity. The output power at 423~nm is 36~mW with a linewidth of 300~kHz. A separate thermal Ca beam is used to keep the absolute frequency in resonance with the atomic transition. The MOT consists of three mutually orthogonal retroreflected beams with diameter of 10~mm and a power of 5~mW in each beam. The frequency of the trapping laser is red detuned by 29~MHz with respect to the resonance. The atoms are directly captured from the low velocity tail of the Boltzmann distribution of a thermal atomic beam. The magnetic field gradient generated by a pair of anti Helmholtz coils is $9\cdot 10^{-3}$~T/cm. A repump laser at 672~nm is used to prevent trap losses via the ${^1}D_1$ state, thereby increasing the density of the atomic ensemble by a factor of four. Absorption imaging of the cloud on the cooling transition was used to determine the radius $r$ and the density $n_0$ of the trapped ensemble leading to typical values of $r=0.3$~mm and $n_0 = 2\cdot 10^{10} \rm{~cm}^{-3}$. The temperature of the ensemble was deduced independently from the ballistic expansion of the cloud and from the Doppler width of the excitation spectrum on the narrow ${^1}$S$_0-{^3}$P$_1$ intercombination line ($\Gamma = 2000$ s$^{-1}$), yielding $2~\textrm{mK}<T<3~\textrm{mK}$. With the repump laser turned on the lifetime of the trap is around 300~ms.

The light of the photoassociation laser is generated by a {Stilben~3} dye laser pumped by a {3~W} Ar$^+$ UV laser. The output power is 100~mW at 423~nm with a linewidth of 1~MHz. The laser is frequency stabilized to a tunable high finesse cavity in vacuum. A tunable helium neon laser at 633~nm is stabilized to the same cavity and its beat frequency with an iodine stabilized helium neon laser is measured. By keeping the beat frequency constant the frequency of the photoassociation laser can be held fixed with an accuracy of $\pm 2$~MHz. Furthermore, with this scheme it is possible to scan the frequency of the photoassociation laser by approx. $1.5$~GHz with the same accuracy. The detuning $\Delta f=f_{\rm Ca}-f_{\rm Laser}$ of the photoassociation laser with respect to the atomic resonance ${^1}$S$_0-{^1}$P$_1$ was determined by measuring its beat frequency with the frequency stabilized trapping laser for detunings up to 20~GHz. For larger detunings the frequency of the photoassociation laser is determined by means of a wavemeter which allows absolute frequency measurements with an accuracy better than $0.2$~MHz. With the transition frequency of the atomic resonance determined the same way the detuning can be calculated.

The rate of photoassociation was measured by recording the increase of trap losses due to photoassociation which manifest themselves in a reduced trap fluorescence. The state of the photoassociated Ca dimer may undergo a state change to a molecular state dissociating into atoms which are no longer trapped (state changing collision), or the dissociative decay of the dimer may lead to a gain in kinetic energy high enough to allow the atoms of the pair to escape from the trap (radiative escape). The photoassociation laser was switched on and off by means of an acousto optical modulator (AOM) for time intervals of 500~ms which must be considerably larger than the trap lifetime of 300~ms to allow the trap to reach a new steady state. The intensity of the photoassociation laser was around 400~mW/cm$^2$. Before and after each interaction period with the photoassociation laser the 423~nm trap fluorescence was detected by an avalanche photodiode. The collisional loss fraction is the ratio of the reduced signal due to photoassociation and the signal obtained without photoassociation. The trapping laser itself stayed on during the whole time. Due to intensity fluctuations of the trap fluorescence of 1~\% to 2~\% averaging is done over at least 30 s to increase the signal-to-noise ratio. The collisional loss fraction is directly proportional to the atomic density of the ensemble and, as long as saturation effects can be neglected, proportional to the intensity of the photoassociation laser.

We measured the collisional trap loss for detunings of the photoassociation laser ranging from $0.6$~GHz to 68~GHz below the atomic ${^1}$S$_0-{^1}$P$_1$ transition. If counting the vibrational levels $v'$ from the asymptote starting at $v'=1$ this frequency interval corresponds to $34\leq v'\leq 72$. To smaller $v'$ we were limited by the resonant interaction of the photoassociation beam with single atoms, pushing them out of the trap by radiation pressure. For higher $v'$ the measurements were limited by decreasing signal strength.

\section{Theoretical Description}
The theoretical description of photoassociation spectra consists of two parts: the precise determination of the bound states close to the dissociation limit and the modelling of bound-free transition together with its detection by trap loss.
To describe the asymptotic level structure for a diatomic molecule like Ca$_2$ we use the accumulated phase approximation which was described by VerHaar et al \cite{moe95}. The potential is given by the long range branch according to the atomic states at the asymptote. For Ca$_2$ at the excited asymptote $^1$S$_0 + ^1$P$_1$ the long range behaviour is dominated by the resonant dipole-dipole interaction and the centrifugal potential term for the rotational state $J$ which should have the correct behaviour for dissociating into an atom pair with non-vanishing atomic angular momentum; for the molecular state $^1\Sigma^+_u$ it becomes:
\begin{equation}
	U(R)=D-\frac{C_3}{R^3}+\frac{\hbar^2[J(J+1)+2]}{2\mu R^2}
\end{equation}
Here $D$ is the asymptotic atom pair energy and $C_3$ is the dipole-dipole interaction constant which can be calculated from the transition moment of  the $^1$S$_0\rightarrow {^1}$P$_1$ transition with frequency $\omega_a$ or from the natural linewidth $\gamma_a$ (all linewidths in this paper are in energy units) of this transition, giving $C_3=3c^3\gamma_a/2{\omega_a}^3$ for the molecular state $^1\Sigma^+_u$, and $\mu$ is the reduced mass of the molecule. 
Because of applying only the potential branch at large internuclear separations we replace the left boundary condition (small $R$) for a bound state by a fixed phase which accumulates from small internuclear separation up to a large $R$, say $R_0$, which is still small enough compared to the classical outer turning point of the desired bound level. If the binding energy of asymptotic states is small compared to the kinetic energy for the vibrational motion around the minimum of the potential the accumulated phase up to $R_0$ is a very weakly varying function for all such asymptotic levels. Thus a fit of observed levels measured with respect to $D$ can be performed with the above potential branch and a simple power expansion of the phase with the binding energy as expansion coefficient. It is numerically easier to choose a weak variation of $R_0$ for a constant phase, where in our case we programmed a maximum of the wave function at the slowly varying $R_0$.

\begin{figure}
\centerline{\includegraphics[width=8cm]{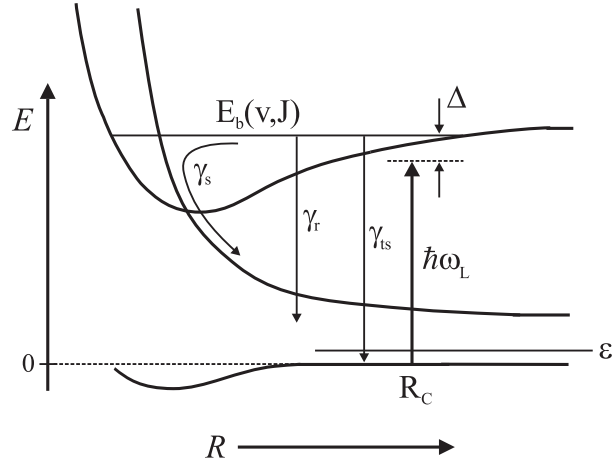}}
 \caption{Schematic potential diagram illustrating the notation used in the text for describing cold collision processes.}
 \label{levels_theo}
 \end{figure}

For modelling the photoassociation profile we apply the theory developed by Bohn and Julienne \cite{boh99}. The coupling of the continuum at the ground state asymptote $^1$S$_0 + ^1$S$_0$ by the laser field to the excited asymptote $^1$S$_0 + ^1$P$_1$ leads to an excitation rate $\Gamma$ and an energy shift $E_1$ of the transition. The above paper shows that the trap loss probability by photoassociation for a collision pair with a kinetic energy $\varepsilon$ can be written in the following form:
\begin{equation}	|S_{0l}|^2=\frac{(\gamma_s+\gamma_r)\Gamma}{[\varepsilon-(\Delta-E_1)]^2+(\frac{\gamma_s+\gamma_r+\gamma_{ts}+\Gamma}{2})^2}	
\end{equation}
$S_{0l}$ is the scattering matrix element of the process of two colliding atoms on the ground state via the bound intermediate state by the light field to the loss state $l$. $\gamma_s$ and $\gamma_r$ are the rates for state changing collisions and the radiative escape, respectively. $\gamma_{ts}$ is the molecular spontaneous decay back to trapped states. $\Delta$ represents the detuning between the excited molecular bound state with energy $E_b(v,J)$ and the laser frequency with respect to the ground state asymptote: $\Delta=E_b-\hbar\omega_L$ (see Fig.~\ref{levels_theo}). $\Delta$ must not be confused with $\Delta f$ which gives the detuning of the photoassociation laser (see Fig.~\ref{levels}a). In the theoretical description we will count the vibrational levels $v$ from the bottom of the potential well. With $v_D$ as the non-integer vibrational `quantum number' derived from the vibrational phase of the hypothetical state exactly at the asymptote one gets the relation $v=v_D-(v_D~\textrm{modulo}~1)-v'+1$. $\Gamma$ and $E_1$ can be evaluated in the reflection approximation, where the scattering wave function of the ground state potential for the scattering energy $\varepsilon$ is only needed at the Condon point of the transition, the above mentioned paper gives the results:
\begin{eqnarray}
	\Gamma=2\pi(V_{0b})^2\frac{\partial E_b}{\partial v}\frac{|f_0(R_C)|^2}{D_C}
\end{eqnarray}
and	
\begin{eqnarray}
	E_1=\pi(V_{0b})^2\frac{\partial E_b}{\partial v}\frac{f_0(R_C)g(R_C)}{D_C}.
\end{eqnarray}
Here the derivative of $E_b$ with respect to $v$ gives the vibrational spacing around the considered bound state, and $f_0$ and $g_0$ are the regular and irregular solution of the scattering function for kinetic energy $\varepsilon$. $R_C$ is the Condon point of the potentials interacting by the laser field at the detuning $\Delta$. $D_C$ is the potential gradient of the excited state at the Condon point, the contribution from the ground state potential plays no rule at large internuclear separation considered in the present case, because this behaves like 1/$R^6$ compared to 1/$R^3$ for the excited state. $V_{0b}$ is the laser interaction which is calculated with the help of the laser intensity $I$, the atomic natural width $\gamma_a$ and the molecular enhancement factor $f_\mathrm{mol}$ which is determined by the molecular symmetry $^1\Sigma^+_u$ and $^1\Sigma^+_g$ of the transition and is to good approximation here 2 (see for example the dipole formula in \cite{mac01}):
$$(V_{0b})^2=\frac{6\pi c^2\gamma_af_\mathrm{mol}}{\omega^3_a}I$$
$\omega_a$ is the atomic transition frequency, here of the $^1$S$_0$ to $^1$P$_1$ transition.
The photoassociation rate $K(\Delta,T)$ at the detuning $\Delta$ from a thermal ensemble at temperature $T$ is then obtained by thermal averaging of the trap loss probability and summing over all contributing partial waves $\ell$ to the bound state $J$:
$$K(\Delta,T,J)=\frac{1}{hQ_T}\int_0^{\infty} d\varepsilon e^{-\frac{\varepsilon}{kT}}\sum_{\ell}(2\ell+1)|S_{0l}(\varepsilon,\ell,J,\Delta,I)|^2$$ 
$S_{0l}$ is shown with all dependencies given in the equations above and $Q_T$ is the partition function for $T$ for the atom pairs with reduced mass $\mu$:
$$Q_T=\left(\frac{2\pi\mu kT}{h^2}\right)^{\frac{3}{2}}$$
For the simulations of photoassociation profiles, especially for the comparison of relative intensities between different vibrational states, one needs a reliable estimate of $\gamma_s$ and $\gamma_r$ at least the relative function with vibrational states. The losses can be either state changing collisions or radiative escape processes, for both we need an inward motion of the photoassociated state. Thus an estimate of the relative probability of the loss will be the vibrational period of the state if one assumes that each crossing of the inner region of the potential leads to an equal amount of loss. The vibrational period can be estimated from the LeRoy-Bernstein formula \cite{ler70} for bound states close to the asymptote: $E_b=D-X(v_D-v)^i$ where $X$ is a constant depending on $C_n$ of the long range behaviour and the reduced mass of the molecule. In the case of the dipole-dipole asymptote $C_3$ is involved and $i$ becomes 6. With this approximation the loss rate $\gamma_s+\gamma_r$ becomes proportional to $6X(v_D-v)^5$; this value was introduced into the simulations of the profiles.
For calculating the eigenenergies with the accumulated phase approximation and the scattering wave function the conventional Numerov method was applied with sufficiently small integration steps, which was always checked by numerical convergence.

\section{Experimental Results}
\begin{figure}
\centerline{\includegraphics[width=13cm]{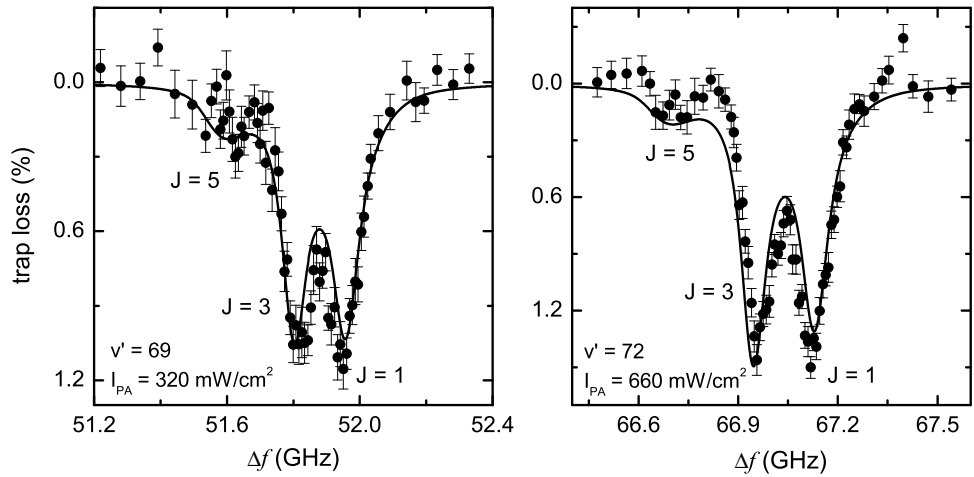}}
 \caption{Vibrational lines $v'=69$ and 72 with three resolved rotational lines corresponding to $J=$ 1, 3 and 5. The solid curve represents the prediction of a full quantum mechanical calculation assuming a scattering length of 85.2~$a_0$.}
 \label{v69_v72}
 \end{figure}

Figure \ref{v69_v72} shows the trap loss spectrum for the vibrational levels $v'=69$ and 72 of the $^1\Sigma_u^+$ state at a detuning $\Delta f$ of the photoassociation laser around 52~GHz and 68~GHz from the atomic resonance. Three rotational lines corresponding to $J=1, 3 \text{ and } 5$ are observed. The sample temperature is about 2~mK, at which the collisional partial waves with $\ell\geq 6$ have too low probability for leading to an observable signal of level $J=7$ of the excited state. The solid curve shows a simulated spectrum. It is obtained by starting with a potential that is known, e.g., from our earlier spectroscopic measurements \cite{zin00}, from which the eigenenergies of the rovibrational levels are calculated with the accumulated phase method. These do not necessarily coincide with the positions of the vibrational levels as measured in the experiments due to the finite temperature, AC Stark shifts and the potential itself. The position of rotational levels with $J$ values of 3 and 5 are shifted up to 120~MHz by these effects. The shift between the observed spectrum and the simulated one is used in an iteration procedure to modify the potential described by $C_3$ and the power expansion of $R_0$. After two to three iterations convergence is achieved. The procedure gives a $C_3=0.5217(45)\cdot 10^3~\textrm{nm}^3\cdot\mathrm{cm}^{-1}$ resulting in an atomic decay rate of $\gamma_a/\hbar = 2.150(19)\cdot 10^8~s^{-1}$. This value is smaller than that reported in \cite{zin00} by three times the error limit. This can be explained by the fact that in \cite{zin00} shifts of the line positions due to temperature have only been accounted for by a global shift of all frequencies of 50~MHz. From our new calculations presented here this is not justified and thus leads to a different value. It was also checked that the retardation effect of the dipole-dipole interaction~\cite{mac01} must be included in the fit. Especially, the levels close to the asymptote $^1$S$_0 + ^1$P$_1$ for $v'<40$ otherwise show non-negligible deviations.

\begin{figure}
\centerline{\includegraphics[width=8cm]{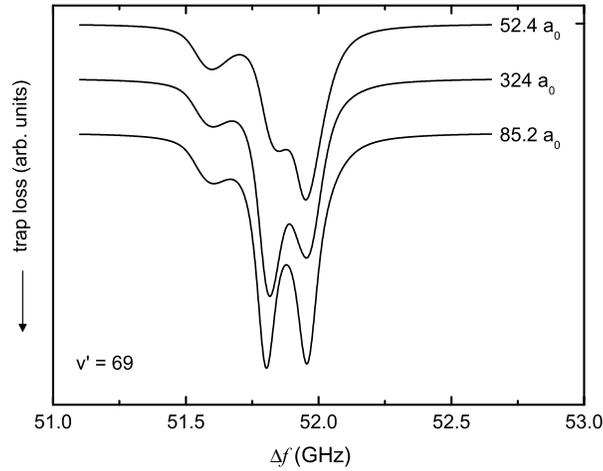}}
 \caption{Simulated spectra for the vibrational level $v'=69$ for 2~mK and 400~mW/cm$^2$ photoassociation laser intensity for three different values of the scattering length. The curves are shifted in the vertical direction for clarity.}
 \label{scat_length}
 \end{figure}

It turns out that the relative height of the components of the rotational substructure depends significantly on the value of the scattering length for the ground state potential. For varying the scattering length in test calculations, one may change the potential slightly at the repulsive branch at small $R$ or around the minimum of the potential. The result are slight shifts of the last bound levels not only for $\ell=0$ but also for $\ell\neq 0$. Here for $\ell=2$ the position of the corresponding shape resonance moves to the asymptote for decreasing scattering length, which results in the variation of the intensity ratio of the peaks for $J=1$ and 3. Fig.~\ref{scat_length} shows the numerical simulation of the vibrational level $v'=69$ for three different scattering lengths of $\mathrm{52.4}~a_0$, $\mathrm{85.2}~a_0$ and 324~$a_0$, respectively, with the Bohr radius $a_0=0.0529\ldots$~nm. For clarity the curves are shifted in the vertical direction. The ratio of the $J=1$ and $J=3$ components and the deepness of the dip between the two components of all measured vibrational lines with resolved rotational substructure ($v'\geq 62$) are best reproduced by the potential with a scattering length of $\mathrm{85.2}~a_0$. The quite different form of the two other curves ($\mathrm{52.4}~a_0$ and 324~$a_0$) suggests that the scattering length lies somewhere between 50~$a_0$ and 300~$a_0$. This is the same order of magnitude of the scattering length as determined by other spectroscopic methods \cite{all02}.

\begin{figure}
\centerline{\includegraphics[width=8cm]{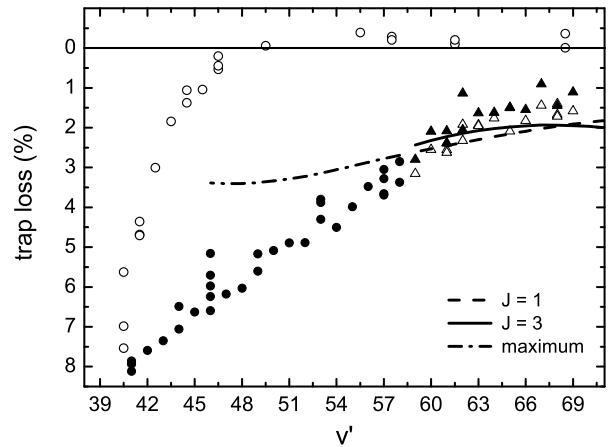}}
 \caption{Measured collisional trap loss as a function of the vibrational quantum number $v'$. For vibrational lines where the rotational substructure overlaps the loss was determined at the expected maximum of the resonance line (full circles). For lines with sufficiently resolved rotational structure the loss was measured for $J=1$ (full triangles) and $J=3$ (open triangles). The curves indicate theoretical predictions with simplified assumptions about decay rates (see text).}
 \label{v_depen}
 \end{figure}

\begin{figure}
\centerline{\includegraphics[width=8cm]{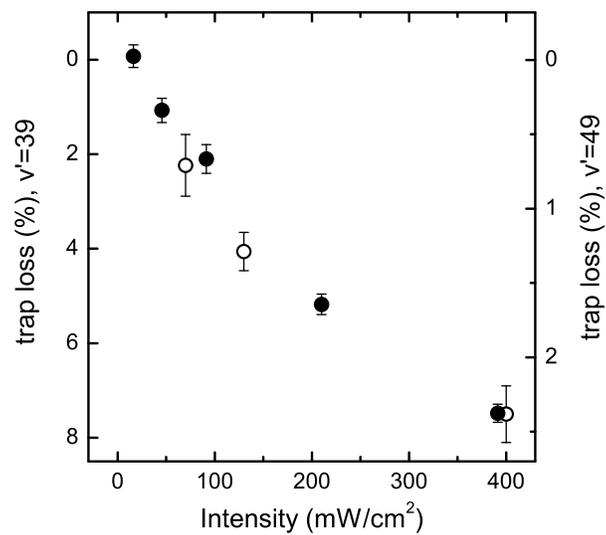}}
 \caption{Dependence of the collisional loss fraction at the peak position of line $v'=39$ (open circles) and line $v'=49$ (full circles) on the photoassociation laser intensity. At the highest intensities used in the experiments saturation effects begin to show up.}
 \label{int}
 \end{figure}

In our previous photoassociation experiments on $^{40}$Ca \cite{zin00} it was not possible to determine the relative intensities of the vibrational lines. This was mainly due to the short trap lifetime of 50~ms which leads to maximum collisional trap loss fractions of 1~\%. In the experiment reported here, the collisional loss fraction at the center of each vibrational line was measured (Fig. \ref{v_depen}). For vibrational lines where the rotational structure is not resolved the collisional loss fraction at the expected maximum of the resonance line was determined (full circles). For large $v'$ the rotational structure is sufficiently resolved and the collisional loss fraction was measured at the position of the $J=1$ (full triangles) and the $J=3$ rotational lines (open triangles). The collisional loss fraction with the photoassociation laser tuned to the gap between two vibrational lines is shown by open circles. For $v'<47$ the vibrational lines begin to overlap and the collisional loss fraction does not drop to zero between two lines. Before and after each measurement session of approximately ten hours the atomic density was determined by taking absorption images on the cooling transition. Due to mechanical drifts in the setup the density decreased over this period of time by up to 20~\%. A linear variation of the atomic density with time was taken into account to normalize the collisional loss fraction to a density of $2\cdot 10^{10}$~cm$^{-3}$. In addition the change in the intensity of the photoassociation laser that showed fluctuations of approx. 10~\% during a scan was corrected. From the measured variation of collisional trap losses with the intensity of the photoassociation laser (Fig. \ref{int}) the fluctuations were corrected to an intensity of 280~mW/cm$^2$.

The curves in Fig.~\ref{v_depen} depict the intensity variation of the trap loss as a function of the vibrational quantum number $v'$ as expected from theory where the functional form of the trap loss in its vibrational dependence is simplified being proportional to $(v_D-v)^5$ from the local vibrational spacing. This approach does not distinguish the contribution by state changing collisions from those by radiative escape. Thus at least a two-parameter ansatz would be necessary for which a functional model does not exist in the literature for describing the vibrational dependence. We were experimenting with different exponents for $(v_D-v)$ instead of 5 and with adding a constant to the loss rate as a second parameter which could model the different weights of the two mechanisms. Such an approach gives a nicer looking graph for the comparison between experiment and theory but not a convincing deeper physical insight. Thus we suppressed it, but we would prefer to indicate clearly with Fig.~\ref{v_depen} the status of our present understanding.

\section{Conclusion}
In conclusion, photoassociation measurements on cold calcium atoms for an extended range of detunings of the photoassociation laser ranging from $0.6$~GHz to 68~GHz below the asymptote were performed and compared with full line profile simulations. We find good agreement between the observed and calculated spectra. The observed intensity ratio of $J=1$ and $J=3$ resonances determines the range of the ground state scattering length between 50~$a_0$ and 300~$a_0$ taking the long range behaviour as determined by spectroscopic experiments~\cite{all02} or by theoretical predictions~\cite{por01}. The measured relative intensities of the vibrational transitions are not well explained. New studies of loss mechanisms are needed or better electronic structure calculations on Ca$_2$ would be desirable to improve on the model calculations which were published by Machholm et al.~\cite{mac01}.

Up to now we only observed the creation of calcium dimers in the $^1\Sigma_u^+$ state. Besides this, the formation of atom pairs in the $^1\Pi_g$ state is allowed due to retardation. Those vibrational levels should show up as very sharp features (a few MHz width) in the spectra \cite{mac01}. It may be possible to get signatures from these states in the experiment when switching to ultracold calcium atoms with a temperature of a few microkelvin.

We gratefully acknowledge O. Allard and A. Pashov for kindly providing us with their data for the potential curves. This work was supported by the Deutsche Forschungs\-gemeinschaft.

\enddocument